\documentclass[%
 reprint,
superscriptaddress,
groupedaddress,
showkeys,
 amsmath,amssymb,
 aps,
floatfix,
]{revtex4-2}
\usepackage{mathrsfs}
\usepackage{amsmath}
\usepackage{bbold}
\usepackage{graphicx}
\usepackage{dcolumn}
\usepackage{bm}
\usepackage{upgreek}

\begin{document}

\preprint{APS/123-QED}

\title{Scattering from Time-modulated Transmission Line Loads: Theory and Experiments in Acoustics}

\author{Matthieu MALL\'EJAC}
\email{matthieu.mallejac@epfl.ch}
\author{Romain FLEURY}%
 \email{romain.fleury@epfl.ch}
 
\affiliation{%
 Laboratory of Wave Engineering, \'Ecole Polytechnique F\'ed\'erale de Lausanne, Switzerland.
}%

\date{\today}

\begin{abstract}
Scattering wave systems that are periodically modulated in time offer many new degrees of freedom to control waves both in spatial and frequency domains. Such systems, albeit linear, do not conserve frequency and require the adaptation of the usual theories and methods. In this paper, we provide a general extension of transmission line or telegraph equations to periodically time-modulated systems. As a by-product of the theory, we obtain a general approach to compute and measure the complete scattering matrix of such systems. Finally, the proposed theory and methods are applied and validated on a concrete practical example in the realm of airborne acoustics: a time-modulated actively controlled loudspeaker membrane terminating a monomode waveguide. Different modulation functions and parameters are tested. The experimental results are compared to both numerical simulation and an analytical model based on a two time-scale method.
\end{abstract}

\maketitle


\section{\label{sec:level1}Introduction}
    Time-varying wave media have attracted a great level of interest over the past few decades and have opened new perspectives in the field of metamaterials by adding a new degree of freedom in wave manipulation and engineering possibilities \cite{galiffi_photonics_2022}. The first studies on wave propagation in time-varying media date back to the mid-nighties, with the work of Morgenthaler \cite{morgenthaler_velocity_1958}, Felsen \textit{et al.} \cite{felsen_wave_1970}, or Fante \cite{fante_transmission_1971} on spatially homogeneous but time-varying dielectric and dispersive media. Scattering from temporal boundary conditions and discontinuities gives rise to intriguing phenomena. The temporal dual of wave scattering on a planar interface between two media cannot, due to causality, involve a reflection to negative times, thus leading to different Fresnel coefficients. Another important contrast between spatial and temporal crystals or metamaterials, \textit{i.e.}, slab of (locally resonant) medium varying periodically in time, is the fact that they can generate not only frequency band gaps, but also wavenumber gaps, corresponding to linearly unstable regimes \cite{koutserimpas_fleury_2018}. 
    
    In particular, systems that are periodically modulated in time, or time-Floquet systems, have the ability to alleviate some of the constraints of simple static media \cite{yin_floquet_2022}, such as the breaking of time-reversal symmetry and of reciprocity \cite{sounas_non-reciprocal_2017,hadad_2020,li_nonreciprocal_2022}. As a result, exciting wave control possibilities open, such as magnet-free circulators and temporal aiming \cite{fleury_subwavelength_2015, pacheco-pena_temporal_2020}, Floquet topological insulators \cite{rechtsman_photonic_2013,fleury_floquet_2016,zhang_superior_2021,zhu_time-periodic_2022}, unidirectional and parametric amplification  \cite{cullen_travelling-wave_1958, koutserimpas_nonreciprocal_2018, koutserimpas_parametric_2018, song_direction_2019, li_nonreciprocal_2019, shen_nonreciprocal_2019, shen_nonreciprocal_2019-1, zhu_tunable_2020, zhu_non-reciprocal_2020, wen_unidirectional_2022}, frequency conversion \cite{lee_linear_2018,wen_unidirectional_2022,zhao_programmable_2019}, holography \cite{bacot_time_2016}, near zero index enabled behaviors (negative refraction, high harmonic generation, time-reversal, broadband, and controllable frequency shift) \cite{vezzoli_optical_2018,koutserimpas_zero_2018,yang_high-harmonic_2019, bruno_negative_2020, zhou_broadband_2020}, or strong non-linear behavior \cite{shan_publisher_2022} allowing the observation of Floquet solitons \cite{mukherjee_observation_2020} or the development of wave-based neuromorphic computing \cite{momeni_electromagnetic_2022}. One of the main characteristics of Floquet metamaterials is the generation of harmonics at integer multiples of the modulation frequency, thus requiring an adaptation of classical experimental or theoretical methods where natural hypotheses such as linearity, reciprocity and frequency conservation are assumed. 
    
    Many efforts have already been made to theoretically describe time-varying systems \cite{pacheco_effective_2020, rizza_nonlocal_2022} and their components \cite{jayathurathnage_time-varying_2021} as well as to extend theoretical tools such as the generalization of Kramer-Kronigs relations \cite{solis_functional_analysis}, of particle's dipolar polarizability \cite{Mirmoosa_2022}, of the transfer matrix methods \cite{li_transfer_2019}, or of the T matrix \cite{garg_modeling_2022}, among others. Transmission line is another key concept fundamental to wave engineering, as it allows for the characterization of wave systems involving the scattering of guided waves. From this theory, one can describe and define the scattering of N-port systems, and easily connect several scatterers together to compose more complex systems. The growing interest for time-varying media requires to extend this theory to multi-harmonic systems, both from theoretical and experimental points of view. 
    
    In this paper, we provide a comprehensive framework to explore the guided-wave scattering of time-modulated loads, from theory to experiments. In particular, the measurement of the complete scattering matrix of such systems, including potential Floquet harmonics, remains an experimental challenge, for which we propose a solution. 
    
    The paper is structured as follows. In a first section, we derive and expose the extended theory of transmission line, scattering and reduced impedance matrices, for time modulated systems. In a second section, we present a method for extracting the complete scattering matrix based on a multiload technique. These two sections are general and can be applied to several domains of wave physics (electrical, acoustic or mechanical transmission lines for example). At last, we take an acoustic example to apply the theory on a concrete case, and demonstrate experimentally our S matrix extraction method: an actively controlled loudspeaker with an assigned input impedance modulated in time. Analytical modeling based on a two time-scale method and numerical simulations are used to confirm the experimental results obtained for different modulation functions and parameters.

\section{\label{sec:theory}Time-modulated transmission line theory}
\begin{figure}
        \centering
        \includegraphics{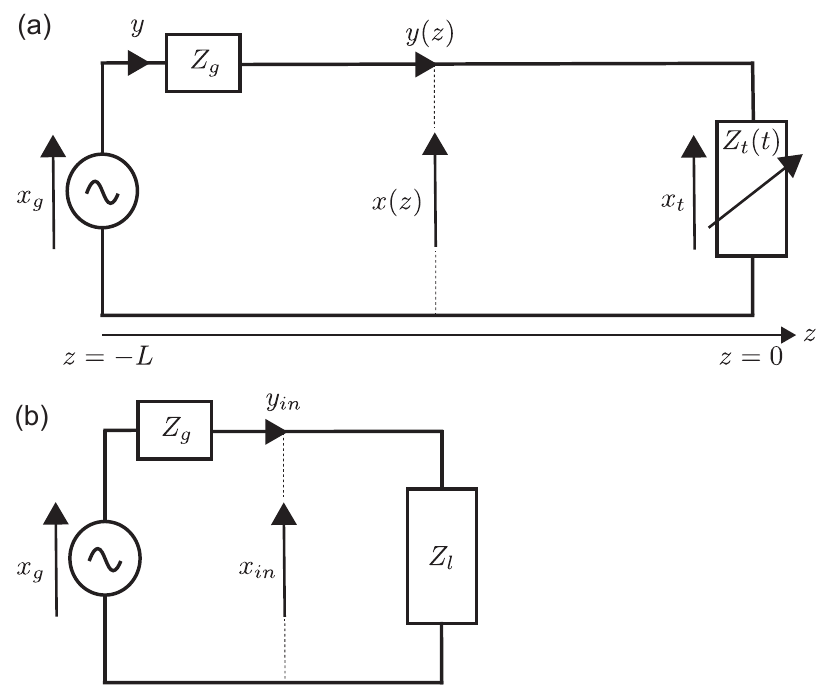}
        \caption{Transmission line circuit terminated by a time-modulated load (a), and reduced impedance circuit (b). }
        \label{fig:schema_circuit}
    \end{figure}

    We consider here a very general one-dimensional transmission line where $x$ and $y$ can represent any quantities such that, in the static case, both are related by the impedance $Z(\omega)$
    \begin{equation}
        x(z,\omega)=Z(z,\omega)\cdot y(z,\omega).
    \end{equation}
    For electrical, acoustic or mechanical circuits, $x$ is respectively the voltage $U$, the pressure $P$, or the force $F$ while $y$ is the current intensity $I$, the particle velocity $V$, or the velocity $V$. The transmission line is terminated by a time-varying load, $Z_t(t)$ as shown in Fig.~\ref{fig:schema_circuit}.

    \subsection{Constant load} 
        We first recall some well-known generalities about transmission lines where the termination load does not vary with time \cite{steer_2019}. In this case, the scalar fields at  position $z$ can be written as the superposition of the incident $x_i$ and reflected $x_r$ waves. Noting  $Z_0$ the characteristic impedance of the line, one has 
        \begin{align}
            x(z,\omega) &= x_i(z,\omega)+x_r(z,\omega),\label{eq:x_extension1}\\
            y(z,\omega) &= \frac{x_i(z,\omega)-x_r(z,\omega)}{Z_0},
            \label{eq:x_extension2}
        \end{align}
        which can be related by a scalar scattering coefficient at $z=0$
        \begin{equation}
            x_r(z=0,\omega) = S_t \cdot x_i(z=0,\omega)
            \label{eq:x_extension3}
        \end{equation}
        \textit{i.e.}, the complex reflection coefficient $S_t=R$.
        Alternatively, the scalar impedance of the load can be related to the reflection coefficient as 
        \begin{equation}
            S_t = R = \frac{Z_t/Z_0-1}{Z_t/Z_0+1},   
            \label{eq:scat}
        \end{equation}
        or, equivalently, 
        \begin{equation}
            Z_t=Z(z=0)=Z_0\frac{1+R}{1-R}
            \label{eq:imp}
        \end{equation}
        
    \subsection{Periodically time-modulated load}
         If the load is now modulated periodically in time with a circular frequency $\omega_m$, Floquet harmonics at $\omega\pm n\omega_m$ ($n\in \mathbb{N}$) are generated around the excitation frequency $\omega$ and will be reflected from the load.  Moreover, since in the general case the source is not impedance matched, the multi-harmonic reflection from the load will also be reflected back from the generator side. 
         Both the incident and the reflected waves can therefore be developed in Fourier series. Since the amplitudes of the harmonics must decrease at large $n$, we can always truncate their summation to a given harmonic order $N$, approximating the signals as follows,
        \begin{align}
            x(z,t) &= \sum_{n=-N}^{N} \mathbf{x}[n](z)\textrm{e}^{\textrm{i}(\omega+n\omega_m)t}, \\ 
            y(z,t) &= \sum_{n=-N}^{N} \mathbf{y}[n](z)\textrm{e}^{\textrm{i}(\omega+n\omega_m)t}, 
        \end{align}
        where $\mathbf{x}[n](z)$ and $\mathbf{y}[n](z)$ are the $n^{\textrm{th}}$ element of the complex amplitude vectors $\mathbf{x}(z)$ and $\mathbf{y}(z)$ of length $2N+1$, defined at each position $z$.

        As a generalization of eqs.~\eqref{eq:x_extension1}, \eqref{eq:x_extension2}, and \eqref{eq:x_extension3}, these complex amplitudes can be expressed as 
        \begin{align}
           \mathbf{x}(z)& = \mathbf{x_i}(z)+\mathbf{x_r}(z)=\left[\mathbb{1}+\mathbf{S}(z)\right]\cdot\mathbf{x_i}(z),\label{eq:xmat}\\
           \mathbf{y}(z)& = \frac{\mathbf{x_i}(z)-\mathbf{x_r}(z)}{Z_0}=\left[\mathbb{1}-\mathbf{S}(z)\right]\cdot\frac{\mathbf{x_i}(z)}{Z_0}.\label{eq:ymat}
        \end{align}
        using the matrix generalization of the scattering coefficient,  eq.~\eqref{eq:x_extension3}.

        The incident and reflected complex amplitude vectors can be expressed as the element wise multiplication of a magnitude vector $\mathbf{x_{i,r}^{\textrm{abs}}}$ and a phase vector $\mathbf{d}(z)$ 
        \begin{equation}
            \mathbf{x_{i,r}}(z) = \mathbf{x_{i,r}^{\textrm{abs}}}\odot\mathbf{d}(z),
        \end{equation}
        where the magnitude vector is given by $\mathbf{x_{i,r}^{\textrm{abs}}}=[|x_{i,r}^{(-N)}|,...,|x_{i,r}^{(0)}|,...,|x_{i,r}^{(+N)}|]^T$.
        The phase vector is defined as  $\mathbf{d}(z) = [\mathrm{e}^{\textrm{i}k^{(-N)}z},...,\mathrm{e}^{\textrm{i}k^{(0)}z},...,\mathrm{e}^{\textrm{i}k^{(+N)}z}]^T$, with the wavenumbers $k^{(n)} = (\omega\pm n \omega_m)/c_0$. 
        The wave celerity in the line is $c_0 =\textrm{const}$. The operator $\odot$ is the Hadamard product, \textit{i.e.}, the element wise multiplication operator.

        Due to the multi-harmonic content, eqs.~\eqref{eq:scat} and \eqref{eq:imp} also need to be generalized. What used to be the scalar scattering coefficient and impedance now become $(2N+1)$ by $(2N+1)$ matrices, related by
        \begin{align}
           \mathbf{Z_t}&=\mathbf{Z}(z=0) = Z_0\left[\mathbb{1}+\mathbf{S_t}\right]\cdot\left[\mathbb{1}-\mathbf{S_t}\right]^{-1}, \label{eq:Zmatrix}\\
           \mathbf{S_t}&=\left[\mathbb{1}+\frac{\mathbf{Z_t}}{Z_0}\right]^{-1}\cdot\left[\frac{\mathbf{Z_t}}{Z_0}-\mathbb{1}\right]\label{eq:Smatrix}.
        \end{align}
        These relations can be easily derived by introducing eqs.~\eqref{eq:xmat} and \eqref{eq:ymat}, in the matrix extended definition of the impedance and scattering coefficient at the load position
        \begin{align}
        \mathbf{x}(z=0)&=\mathbf{Z_t}\cdot\mathbf{y}(z=0),\label{eq:generalizedZ}\\
        \mathbf{x_r}(z=0)& = \mathbf{S_t}\cdot \mathbf{x_i}(z=0). \label{eq:generalizedS}
        \end{align}
        
        We can then determine the scattering and impedance matrices at any $z$ position along the transmission line from the ones at the termination. The reduced impedance can be expressed as
        \begin{align}
                \mathbf{Z}(z)=&\resizebox{.8\hsize}{!}{$Z_0 \left[\mathbb{1}+ \left[\mathbb{1}+\frac{\mathbf{Z_t}}{Z_0}\right]^{-1}\cdot\left[\frac{\mathbf{Z_t}}{Z_0}-\mathbb{1}\right]\odot \left(\mathbf{d}(z)\cdot\mathbf{d}(z)^T\right)\right.$}\nonumber&\\
                &\cdot \resizebox{.74\hsize}{!}{$\left[\mathbb{1} - \left[\mathbb{1}+\frac{\mathbf{Z_t}}{Z_0}\right]^{-1}\cdot\left[\frac{\mathbf{Z_t}}{Z_0}-\mathbb{1}\right]\odot \left(\mathbf{d}(z)\cdot\mathbf{d}(z)^T\right)\right]^{-1}$},
            \label{eq:reducedZ}
        \end{align}
        and the scattering matrix reduced to the $z$ position is given by $\mathbf{S}(z)=\mathbf{S_t}\odot \left(\mathbf{d}(z)\cdot\mathbf{d}(z)^T\right)$.
        
        Finally, we can compute the average power along the transmission line 
        \begin{align}
           \mathscr{P}(z)=\frac{1}{2}\Re \{ \mathbf{x} \cdot \mathbf{y}^* \} = &\left[\mathbb{1}+\mathbf{S}(z)\right]\cdot\left[\mathbf{x_i^{\textrm{abs}}}\odot\mathbf{d}^*(z)\right]\nonumber
           \\
           &\cdot\left[\mathbb{1}-\mathbf{S}(z)\right]^*\cdot\frac{\mathbf{x_i^{\textrm{abs}}}\odot\mathbf{d}(z)}{Z_0}.
        \end{align}
        
        In terms of incident $\mathscr{P}_i$ and reflected $\mathscr{P}_r$ power, we have
        \begin{equation}
            \mathscr{P}(z) = \mathscr{P}_i(z)+\mathscr{P}_r(z)=\frac{1}{2Z_0}|\mathbf{x_i}|^2-\frac{1}{2Z_0}|\mathbf{S}(z)\cdot\mathbf{x_i}|^2.
        \end{equation}
        According to Fig.~\ref{fig:schema_circuit}(b), the incident wave can also be defined with respect to the generator voltage, pressure or force through the load impedance reduced to the generator position
        \begin{align}
            \mathbf{x_{in}}&=\mathbf{Z_{in}}\cdot\left[\mathbf{Z_s}+\mathbf{Z_{in}}\right]^{-1}\cdot\mathbf{x_g}\nonumber\\
            &=\left[\mathbb{1}+\mathbf{S}(-L)\right]\cdot\mathbf{x_i^{\textrm{abs}}}\odot \mathbf{d}^*(-L),
        \end{align}
        with the input impedance $\mathbf{Z_{in}}$ being defined from  eq.~\eqref{eq:reducedZ}
        \begin{equation}
            \mathbf{Z_{in}} = \mathbf{Z}(-L).
        \end{equation}
    These equations provide, from the measured time signals $x(t)$ and/or $y(t)$ developed in Fourier series, the total field at each Floquet harmonic, $\mathbf{x}$ and $\mathbf{y}$,  and allow to fully characterize a time-modulated load in terms of scattering, impedance and power.
    
\section{\label{sec:extract}Extraction of the scattering matrix}
    Extracting the complete scattering matrix of a time-modulated system remains a challenge due to the multi-harmonic content. One needs to measure $2N+1$ linearly independent data sets from which one can extract any of the elements of the $\mathbf{S_t}$ matrix.
    
    Here, we propose a general methodology based on multi-load techniques \cite{Sitel_Multiload_2006} to measure the $(2N+1)$ by $(2N+1)$ scattering matrix while limiting the number of sensors required. It is worth noting here that another method would consist in multiplying the number of sensors in the system.
    
    Two extraction methods can be employed. First, $\mathbf{S_t}$ can be extracted by probing $x(t)$ at two positions in the transmission line, which allows differentiating the incident $\mathbf{x_i}$ and reflected $\mathbf{x_r}$ fields for each harmonic that are related to $\mathbf{S_t}$ through eq.~\eqref{eq:generalizedS}. For the second method, we can use eq.~\eqref{eq:Zmatrix} to obtain $\mathbf{S_t}$ from the $\mathbf{Z_t}$ matrix that relates the harmonic content of the $x(t)$ and $y(t)$ fields measured at the load position (eq.~\eqref{eq:generalizedZ}).

    \subsection{Multiload technique}\label{sec:multiload}
        Both methods involve inverting equations with vectors. To do this, we expand the vectors $\mathbf{x_i}$, $\mathbf{x_r}$ or $\mathbf{x}$ and $\mathbf{y}$ into square matrices, multiplying the number of measurements, thus multiplying the number of equations, to solve the $(2N+1)^2$ unknowns.
        In order to invert these matrices, the equations must be linearly independent in the frequency range of interest.  We can then vary the length of the transmission line between the load to be characterized and the probe position for each measurement. By doing so, we can change the phase of the reflected field at the microphone position. The length should be chosen so that each configuration gives independent data sets over the frequency range for $\mathbf{x_i}$ and $\mathbf{x_r}$,  or $\mathbf{x}(z=0)$ and $\mathbf{y}(z=0)$. In addition, these methods assume plane wave propagation. The excitation frequency as well as all generated Floquet harmonics must be lower than the waveguide cutoff frequency.

        Equations \eqref{eq:generalizedS} and \eqref{eq:generalizedZ} can then be rewritten as
        \begin{equation}
        \mathbf{S_t}  =
            \begin{bmatrix}
                x_{r_{L_1}}^{-N} & \cdots & x_{r_{L_{2N+1}}}^{-N}\\
                \cdot & \cdots &\cdot \\
                \cdot & \cdots &\cdot \\
                x_{r_{L_1}}^{0} & \cdots & x_{r_{L_{2N+1}}}^{0}\\
                \cdot & \cdots & \cdot \\
                \cdot & \cdots& \cdot \\
                x_{r_{L_1}}^{+N} & \cdots & x_{r_{L_{2N+1}}}^{+N}\\
            \end{bmatrix}\cdot
            \begin{bmatrix}
                x_{i_{L_1}}^{-N} & \cdots & x_{i_{L_{2N+1}}}^{-N}\\
                \cdot & \cdots &\cdot \\
                \cdot & \cdots &\cdot \\
                x_{i_{L_1}}^{0} & \cdots & x_{i_{L_{2N+1}}}^{0}\\
                \cdot & \cdots & \cdot \\
                \cdot & \cdots& \cdot \\
                x_{i_{L_1}}^{+N} & \cdots & x_{i_{L_{2N+1}}}^{+N}\\
            \end{bmatrix}^{-1},
            \label{eq:Z_from_pipr} 
        \end{equation}
        \begin{equation}
        \mathbf{Z_t}=
            \begin{bmatrix}
                x_{{L_1}}^{-N} & \cdots & x_{{L_{2N+1}}}^{-N}\\
                \cdot & \cdots &\cdot \\
                \cdot & \cdots &\cdot \\
                x_{{L_1}}^{0} & \cdots & x_{{L_{2N+1}}}^{0}\\
                \cdot & \cdots & \cdot \\
                \cdot & \cdots& \cdot \\
                x_{{L_1}}^{+N} & \cdots & x_{{L_{2N+1}}}^{+N}\\
            \end{bmatrix}_{z=0}\cdot
            \begin{bmatrix}
                y_{{L_1}}^{-N} & \cdots & y_{{L_{2N+1}}}^{-N}\\
                \cdot & \cdots &\cdot \\
                \cdot & \cdots &\cdot \\
                y_{{L_1}}^{0} & \cdots & y_{{L_{2N+1}}}^{0}\\
                \cdot & \cdots & \cdot \\
                \cdot & \cdots& \cdot \\
                y_{{L_1}}^{+N} & \cdots & y_{{L_{2N+1}}}^{+N}\\
            \end{bmatrix}_{z=0}^{-1}.
            \label{eq:Z_from_p_V}
        \end{equation}
        
        The inversion of these square matrices is sensitive to the independence of the data sets, \textit{i.e.}, to the choice of the load lengths. Computing the condition number of the matrix to be inverted allows to select correctly the adequate lengths.
        Initially, the different lengths can be chosen so that, for the mid-range frequency, each produces a phase shift linearly distributed in the range $[0\textrm{ }\pi]$.
        
        To increase the accuracy for each frequency in the band, one can also overdetermine the system by increasing the number of configurations to be measured to $M$, \textit{i.e.}, $M$ of transmission line lengths. One can then solve the pseudo-inverse of a matrix $(2N+1)$ by $(M)$ with a least mean squares procedure, thus minimizing the uncertainties in the inversion process due to matrices that may be close to singularity at some frequencies, \textit{i.e.}, data sets that are not sufficiently independent.  
    
    \subsection{$\mathbf{S}$ matrix from $\mathbf{x}_i$ and $\mathbf{x}_r$ discrimination}
        For each of the $2N+1$ charges, the incident $x_i^{(\pm n)}$ and reflected $x_r^{(\pm n)}$ complex amplitudes can be discriminated from the total field $x^{(\pm n)}(z)$ at each harmonic using two carefully calibrated probes positioned at $z_1$ and $z_2$,
        \begin{equation}
           \resizebox{.88\hsize}{!}{$\begin{bmatrix}
                x_{i}^{(\pm n)} & x_{r}^{(\pm n)}\\
            \end{bmatrix}
            =
             \begin{bmatrix}
                x^{(\pm n)}(z_1) & x^{(\pm n)}(z_2)\\
            \end{bmatrix}\cdot
            \begin{bmatrix}
                \textrm{e}^{-\textrm{i}k^{(\pm n)}z_1} & \textrm{e}^{-\textrm{i}k^{(\pm n)}z_2}\\
                \textrm{e}^{\textrm{i}k^{(\pm n)}z_1} & \textrm{e}^{\textrm{i}k^{(\pm n)}z_2}\\
            \end{bmatrix}^{-1}.$}
            \label{eq:incident_ref_mat}
        \end{equation}    
         One can then access $\mathbf{S_t}$ from the $2N+1$ vectors $\mathbf{x_r}$ and $\mathbf{x_i}$ concatenated in matrices, as detailed in eq.~\eqref{eq:Z_from_pipr}.
        This approach assumes linearity along the transmission line, and requires that individual harmonics do not interact with each other during propagation. The sole interactions should occur at the load.
    
    \subsection{$\mathbf{S}$ matrix from $\mathbf{Z}$ matrix }
        The impedance matrix can be obtained directly by simultaneously probing, at the load position, the total fields $x^{(\pm n)}$ and $y^{(\pm n)}$ at each harmonic and for the $2N+1$ charges
        \begin{equation}
        \mathbf{x}(z=0) =
            \begin{bmatrix}
                x_{z=0}^{(-N)}\\ \cdot \\ \cdot \\ x_{z=0}^{(0)} \\ \cdot \\ \cdot \\ x_{z=0}^{(+N)} \\
            \end{bmatrix}
            =
             \mathbf{Z_t}  \cdot
             \begin{bmatrix}
                y_{z=0}^{(-N)} \\ \cdot \\ \cdot \\ y_{z=0}^{(0)} \\ \cdot \\ \cdot \\ y_{z=0}^{(+N)} \\
            \end{bmatrix}
            =
             \mathbf{Z}_{t} \cdot \mathbf{y}(z=0).
            \label{eq:impedance_mat}
        \end{equation}  
        The scattering matrix can then be deduced from the impedance matrix by simply solving eq.~\eqref{eq:Zmatrix}.

\section{Application to a time-modulated actively controlled loudspeaker}
    We now apply the general theory and the extraction procedure of the scattering matrix to an acoustic example. 

    \subsection{Experimental set-up}
\begin{figure}
        \centering
        \includegraphics{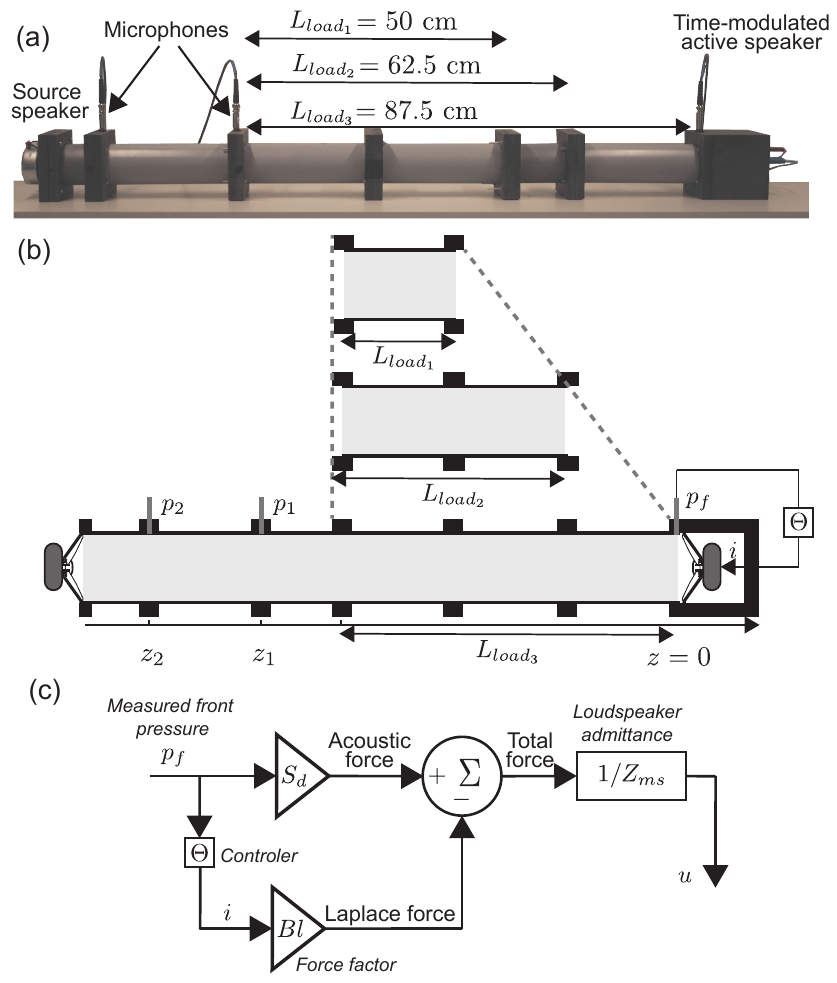}
        \caption{Photograph (a) and schematic (b) of the experimental set-up used to measure the entire scattering matrix of a time-modulated acoustic system (a), block diagram of the active control strategy (b).}
        \label{fig:exp_setup}
    \end{figure}

    We consider a one dimensional circular cross-section acoustic waveguide of diameter $d=7.18$ cm terminated by an actively controlled loudspeaker enclosed in a cavity of volume $V_b = 1081.6$ cm$^3$ as depicted in Fig.~\ref{fig:exp_setup}(a,b). The waveguide is instrumented with two microphones to measure the incident and reflected pressures, and is excited from the left by a monochromatic wave of circular frequency $\omega$.
    
    With active control, the characteristics of a resonator, such as the electrodynamic loudspeaker, can differ completely from its natural properties, e.g., modified resonant frequency, stiffness, impedance \cite{koutserimpas_active_2019}, and non-linearity, paving the way for a plethora of applications such as nonreciprocal behavior \cite{penelet_broadband_2021}, gain and loss control \cite{fleury_invisible_2015, rivet_constant-pressure_2018} or enhanced broadband absorption \cite{furstoss_surface_1997, collet_active_2009, lissek_electroacoustic_2011, rivet_broadband_2017, guo_improving_2020,guo_pid-like_2022} among others. 
    
    Here, we periodically modulate in time the impedance of the loudspeaker, so that it responds with a target impedance $Z_{\textrm{targ}}(t)$. The control part illustrated in Fig.~\ref{fig:exp_setup}(c) and performed using an FPGA based Speedgoat Performance Real-Time controller (I/O 131), consists first in measuring the pressure $p_f$ in front of the loudspeaker and then applying a feedback loop that assigns a given current $i(t)$ to the speaker, based on a given control law $\Theta$
        \begin{equation}
            i = \Theta p_f  = \frac{S_d}{Bl}\left(1-\frac{Z_{ms}(\omega_c)}{Z_{\textrm{targ}}(t)}\right)p_f,
            \label{eq:control_law}
        \end{equation}
        where $S_d$ is the speaker effective cross-section, $Bl$ is the Force factor, and $Z_{ms}$ is the specific impedance of the loudspeaker.

    In the following, we will first consider a narrowband control, allowing the response of the resonator to time modulation to be carefully studied while decoupling the effect of the control at other frequencies. Apart from the narrow frequency range where the modulation occurs, no Floquet harmonics are generated. Only the middle column of the scattering matrix is thus meaningful, i.e., the reflection at the different harmonics for an incidence at the excitation frequency.
    Finally, we will demonstrate the experimental extraction of the full scattering matrix using a broadband control law, allowing the generation of harmonics for any excitation frequency and making the characterization of the complete matrix relevant.

    \subsection{Narrow band control}
        To limit the control only over a given bandwidth $B_c$ around the control frequency $f_c$, a complex envelope technique based on a second order Bessel function is adopted. More details on this technique and the control efficiency can be found in \cite{koutserimpas_active_2019}.
    
\begin{figure*}
            \centering
            \includegraphics{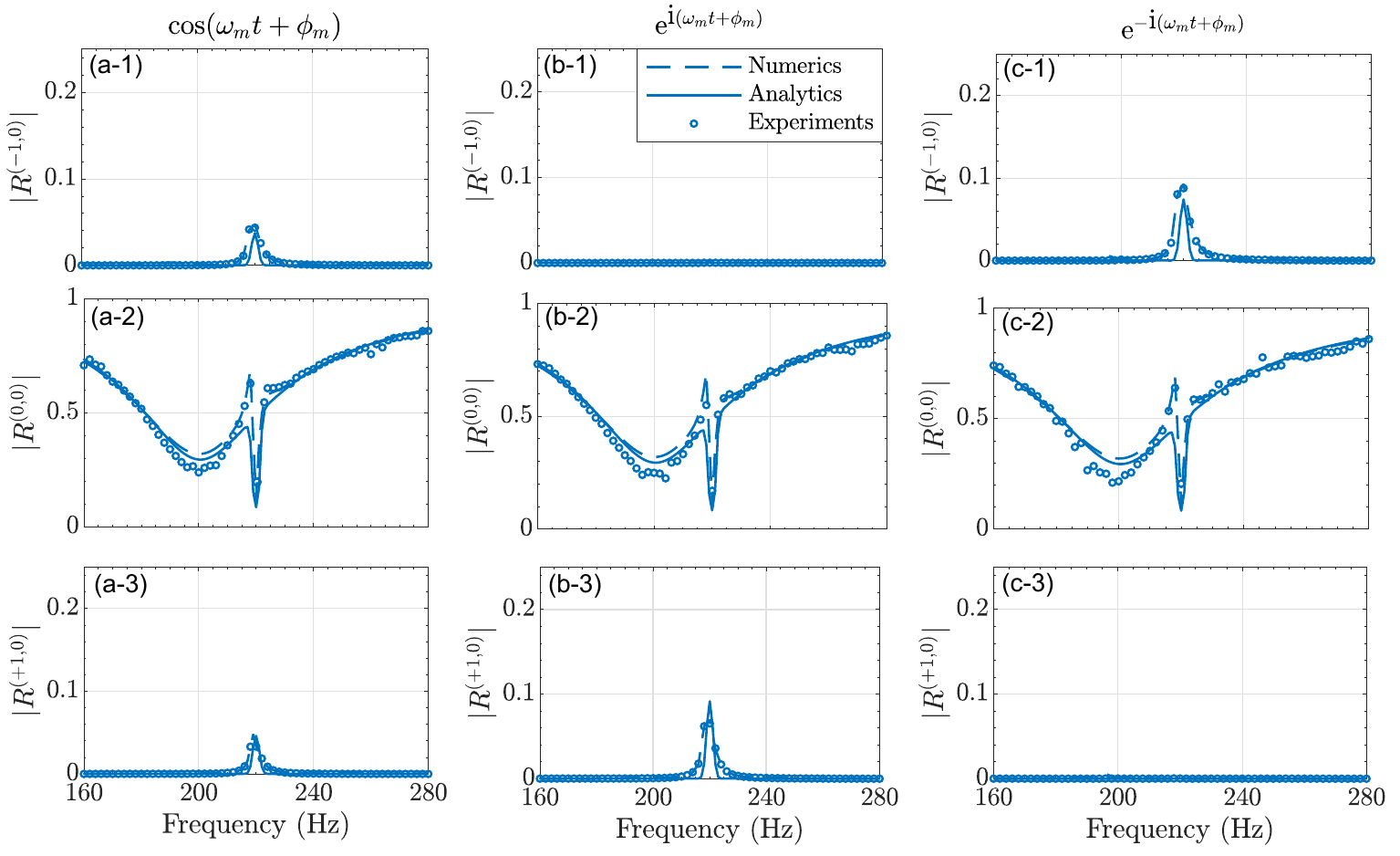}
            \caption{\textbf{Effect of the modulation function:} second column of the full scattering matrix for a cosine modulation (a), positive circular modulation (b), and negative circular modulation (c). The first line, second, and third lines corresponds respectively to $|\mathbf{S}_{12}|=|R^{(-1,0)}|$, $|\mathbf{S}_{22}|=|R^{(0,0)}|$ and $|\mathbf{S}_{32}|=|R^{(+1,0)}|$ respectively. Analytical, numerical, and experimental results are given respectively by the solid and dashed lines and the symbols. }
            \label{fig:S_diff_func}
        \end{figure*}

        Two different periodic modulation functions are investigated in Fig.~\ref{fig:S_diff_func}(a,b,c), a cosine and a positive/negative circular modulation respectively such that
        \begin{align}
             &Z_{targ}(t)= \Tilde{Z}_t\left(1+A_m\cos(\omega_m t+\phi_m)\right),\\
           \textrm{or } &Z_{targ}(t)=\Tilde{Z}_t\left(1+A_m\exp(\pm \textrm{i}(\omega_m t+\phi_m))\right),
        \end{align}
        where $A_m$, $\phi_m$, and $\omega_m = 2\pi f_m$ are the modulation depth, phase, and circular frequency respectively, and $\Tilde{Z}_t =Z_t/Z_0$ is the amplitude of the target normalized impedance.

        \vspace{0.5cm}
        To characterize the effect of the modulation functions, the second column of the scattering matrix is extracted experimentally following the procedure detailed in Section~\ref{sec:extract}. 
        In the following examples, the control characteristics are fixed as follows: $f_c = 220$ Hz, $B_c = 2$ Hz, $Z_{t} = 0.8$, $A_m = 0.2$, and $f_m = 50$ Hz. A prior pressure measurement along the transmission line ended by the time-modulated load, showed that only the first Floquet harmonic is measurable in the system. Thus, the truncation in the Fourier series is set as $N=1$.

        Due to the control, the reflection at $\omega$ for an incidence at $\omega$ exhibits that of the natural loudspeaker, except in the control bandwidth around $f_c = 220$ Hz, where the reflection reaches the value given by the target impedance $|R|=|(\Tilde{Z}_t-1)/(1+\Tilde{Z}_t)|$, as shown in Figs.~\ref{fig:S_diff_func}(a,b,c-2).
        This is also the case for a time invariant control, except that all the extra-diagonal terms of the matrix $\mathbf{S_t}$ are in this case null. 
        Non-zero off-diagonal terms appear only when time modulation is enabled. It generates, in addition to the change in the reflection $|R^{(0,0)}|$ in the control range, some reflection at the $\pm 1$ Floquet harmonics. For example, two reflection peaks centered on $f_c$ are now visible in Fig.~\ref{fig:S_diff_func}(a-1) and (a-3), corresponding respectively to a reflection at $\omega \pm \omega_m$ in response to an incidence at $\omega$, \textit{i.e.}, $|\mathbf{S}_{32}|=|R^{(+1,0)}|$ and $|\mathbf{S}_{12}|=|R^{(-1,0)}|$.
    
        Cosine modulation then generates both positive and negative Floquet harmonics (Figs.~\ref{fig:S_diff_func}(a-1,3)) while a complex function, e.g. a positive (resp. negative) complex exponential, generates only positive (resp. negative)  Floquet harmonics as evidenced in Fig.~\ref{fig:S_diff_func}(a-1) (resp. Fig.~\ref{fig:S_diff_func}(c-3)). 

        We compare the experimental  results (circle symbols) with numerical simulation  based on a Finite Difference Time Domain approach, FDTD, (dashed lines) and an analytical model based on a two-time scale approach (solid lines). More details on the analytical and numerical modeling can be found in Appendix~\ref{app:model}. It is noteworthy here that these three methods require prior and accurate characterization of the loudspeaker and evaluation of its mechanical parameters (see Appendix~\ref{app:model}). 

        We can note a slight discrepancy just before the control frequency between the analytical model and the numerical and experimental results. This can be explained by the way the control bandwidth is applied. Indeed, both numerically and experimentally, a complex envelope technique involving 2nd order Bessel filtering is used, whereas the analytical modeling involves only a generalized normal distribution window (see Appendix~\ref{app:model}).
        Nevertheless, it is worth noting that the analytical model matches well with the expected numerical and experimental reflection value at $f_c$, both at the fundamental and at the first positive and negative Floquet harmonics, thus validating our analytical modeling and the extraction procedure.

        A detailed analysis of the effect of the modulation depth, target impedance, and modulation frequency can be found in Appendix~\ref{app:detail}.
             
    \subsection{Broadband control}
    Extracting the full scattering matrix is only relevant if the time modulated load is effective also at the frequency of the harmonics. Then, the generated harmonic, for example at $\omega\pm\omega_m$ generates back a Floquet harmonic that contribute to the reflection at $\omega$, making the measurement more challenging. 
    
    A substantial incident amplitude at $\omega$, and $\omega \pm\omega_m$ is required to measure these backscatter coefficients, which correspond to the first and third columns of the scattering matrix in our case.
    Floquet harmonics then need to be generated by the time-modulated load for any incident frequency. In other words, the control has to be broadband and has to generate large amplitude harmonics.

    To do so, we change the control law applied to the system and modulate no longer the magnitude of the load impedance but it's compliance $C_{ms}$, \textit{i.e.}, it's resonance frequency, 
    \begin{equation}
        Z_{\textrm{targ}}(\omega,t) = Z_{ms}(\omega)+(\textrm{i}\omega C_{ms})^{-1}A_m\cos(\omega_m t+\phi_m).
    \end{equation}
    
    This new broadband control law still generates mostly one positive and one negative Floquet harmonics, but its stronger effect increases their amplitude, allowing the extraction of the complete $\mathbf{S_t}$ matrix.

       \begin{figure*}
        \centering
    \includegraphics{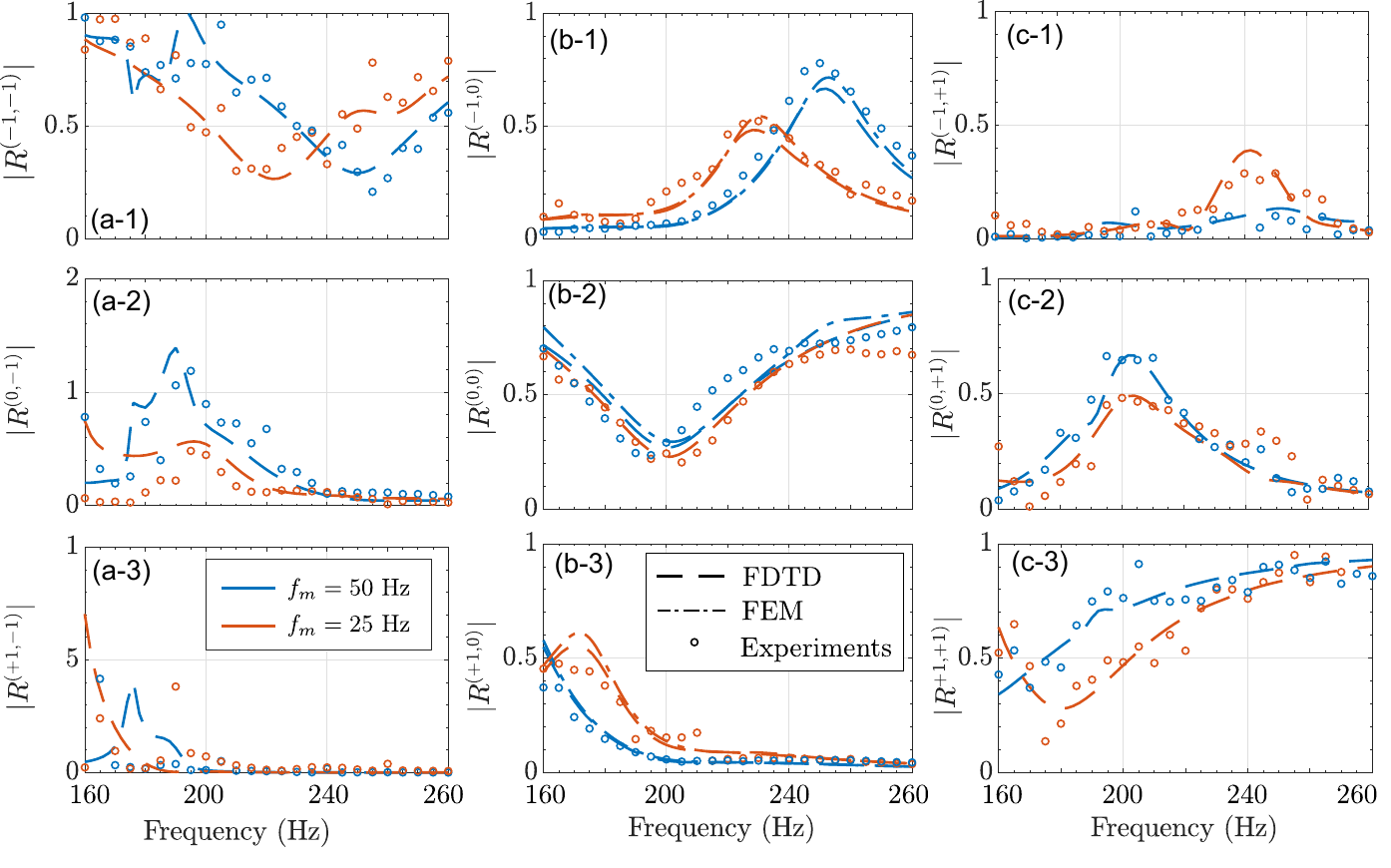}
        \caption{\textbf{Full scattering matrix of the time modulated loudspeaker} for a modulation $f_m = 50$ Hz (blue) or $f_m = 25$ Hz (red), $\phi_m = 0$, and $Z_{\textrm{targ}} = Z_{ms}(\omega)+0.2 (\textrm{i}\omega C_{ms})^{-1}\cos(\omega_m t+\phi_m)$. Reflection coefficient towards $\omega-\omega_m$ (1), towards $\omega$ (2), and towards $\omega+\omega_m$ (3) for an incidence at $\omega-\omega_m$ (a), at $\omega$ (b), and at $\omega+\omega_m$ (c). FEM (only for the 2nd column) and FDTD simulations are given respectively by the dashed and dashed-dotted lines. Experimental results are given by the circle symbols.}
        \label{fig:full_S}
    \end{figure*}

     Figure~\ref{fig:full_S} shows two complete 3-by-3 scattering matrices obtained for two different modulation frequencies $f_m = 50$ Hz (blue) and $f_m = 25$ Hz (red). The experimental results represented by the circle symbols are compared to the scattering coefficients extracted using the 2-probe multiload technique applied to the FDTD experiment. To complement these two methods, a finite element model based on a Fourier expansion of the target impedance is also developed to validate the second column of the scattering matrix. Only the reflections for an incidence at the excitation frequency, \textit{i.e.}, 2nd column only, is accessible with the FEM model (dashed line, see Appendix~\ref{app:model} for more details).
          
    The first, second and third columns of the scattering matrix, see Figs.~\ref{fig:full_S}(a,b,c), correspond respectively to the reflection from an incidence at $\omega-\omega_m$, $\omega$, and $\omega+\omega_m$, to $\omega-\omega_m$, $\omega$, and $\omega+\omega_m$ for the first, second, and third line elements Figs.~\ref{fig:full_S}(1,2,3).
        
    The reflection at the excitation frequency is again that of the speaker with a drop in reflection at the natural resonant frequency, \textit{i.e.}, 200 Hz for $|\mathbf{S}_{22}|$ in Fig.~\ref{fig:full_S}(b-2). 
    Since the first and third columns refer to what happens at $\omega \pm \omega_m$, the reflection curve and the drop are therefore delocalized to $200$ Hz $\pm f_m$ for $|\mathbf{S}_{33}|$ and $|\mathbf{S}_{11}|$, shown in Fig.~\ref{fig:full_S}(c-3, a-1) respectively. 
        
    The scattering coefficients of these two columns are relying mainly on the incident pressures measured at $\omega\pm\omega_m$ which are uniquely due to the generated harmonics and are thus more sensitive to noise, hence the larger variance of the experimental data. Another source of discrepancies comes from the condition number of the matrix to be inverted, which depends on the chosen lengths of the multi-loads and is optimal only for the centre of the frequency range but not necessarily for all frequencies of the bandwidth.
    The difference in amplitude of the different off-diagonal terms is explained by the frequency dispersion of the speaker.  

    Except slight discrepancies for some frequencies, the overall agreement of the measured data with the simulation of the full experimental set-up (including the dispersion and mechanical damping of the source) demonstrates the ability to measure the full scattering matrix of time-modulated systems. 
    
\section{Conclusions}
    In conclusion, we have studied the effect of a time-modulated load on a typical transmission line, and we have extended the classical theory to include the Floquet harmonic generated by the system. We have tackled the challenge of experimentally characterizing the scattering of such structures,
by implementing a multi-load measurement technique. The characterization of time-modulated building blocks is an essential element for the design of more complex devices like space-time varying metamaterials. The extended transmission line theory and scattering extraction methodology are verified and applied to a one dimensional acoustic transmission line terminated by a time-modulated load. To do so, we periodically modulated  in time the input impedance of an actively controlled loudspeaker. The experimental scattering is confronted with both time-domain numerical simulations and analytical modeling based on a 2 time-scale model of the controlled loudspeaker. The agreement of the three is very good for the different functions and modulation parameters tested. An interesting feature is the possibility to force the generation of either positive and/or negative Floquet harmonics solely by the choice of the modulation function applied in the control law. This unique behavior could be used to improve sound absorption by transferring low-frequency acoustic energy only to the higher harmonics, which can be absorbed more easily..
    In addition, to the extended telegraphers equations, that could also be applied to the characterization of nonlinear load, we also present an analytically well described time-modulated controllable acoustic system which can be used in various application of time-varying phenomena such as non-reciprocal device, acoustic circulators, and non-hermitian systems among others. 

\appendix

\section{\label{app:model}Details on the modelling}
    \subsection{Analytical model - 2 time-scale method}
        The actively controlled loudspeaker follows an integro-differential equation relating the velocity of the loudspeaker diaphragm $v(t)$ to the pressure in front of the loudspeaker $p_f(t)$
        \begin{align}
            \nonumber\left( M_{ms}\textrm{d}^2_{tt} \right. & \left. + R_{ms}\textrm{d}_t+[C_{ms}]^{-1} \right) v(t)  \\  & = S_d \textrm{d}_t p_f(t)-Bl \textrm{d}_t i(t)W(\omega),
            \label{eq:ode_speak}
        \end{align}
        where $R_{ms}$, $M_{ms}$ and $C_{ms}$ are the Thiele and Small characteristics of the loudspeaker, \textit{i.e.}, acoustic resistance, mass and compliance, respectively, and $W(\omega)$ is a frequency generalized normal distribution window centered at $f_c$.

        Inserting eq.~\eqref{eq:control_law} into eq.~\eqref{eq:ode_speak}, imposing the change of variable $\tau = \omega_\infty t$ to obtain a dimensionless time variable, and rearranging the terms, we obtain
        \begin{align}
             &\left(\frac{}{}\textrm{d}^2_{\tau \tau}+\frac{R_{ms}}{M_{ms}\omega_\infty} \textrm{d}_{\tau} + 1\right)v(\tau) =\frac{S_d}{ M_{ms}\omega_\infty} \label{eq:ode_speak32} \\& \nonumber \resizebox{.95\hsize}{!}{$\times\textrm{d}_{\tau}  \left[1-\left(1-\frac{Z_{ms}(\omega_c)}{Z_{t}\left(1+A\cos\left(\frac{\omega_m}{\omega_\infty} \tau+\phi_m\right)\right)}\right)W(\omega)\right] p_f(\tau),$} 
        \end{align}
        with $\omega_\infty^2  = [C_{ms}M_{ms}]^{-1}$ the natural  resonance circular frequency of the loudspeaker.
        
        The system is excited by a source delivering a pressure $p(\tau) = P_{inc}\exp\left(\textrm{i}\omega/\omega_\infty \tau\right)$.
        
        At $\tau = 0$ (initial condition), the driven loudspeaker has zero acceleration $\textrm{d}_\tau v(\tau)_{|\tau = 0}=0$ and has a velocity equal to $S_d v(\tau = 0) = p_f(\tau = 0)/Z_{ms} = P_{inc}/Z_{ms}$.
        
        We assume that the system can be described using two different time scales. One related to the excitation at $\omega$, $T_0 \approx \tau$, and the other related to the slow modulation at $\omega_m$, $T_1 \approx \epsilon \tau$, such that $T_0>>T_1$. We define as a small parameter for the derivations, $\epsilon \approx \omega_m/\omega_\infty << 1$. We also note that the prefactor of the right-hand side of eq.~\eqref{eq:ode_speak32} is of the same order as $\epsilon$, so $S_d/M_{ms}\omega_\infty$ can be replaced by $\epsilon S_d/M_{ms}\omega_m$.
        
        The velocity field can then be extended according to these two scales
        \begin{equation}
            v(\tau) \approx v_{0}(T_0,T_1) + \epsilon v_{1}(T_0,T_1).
            \label{eq:exp_v}
        \end{equation}

        The governing equation can then be rewritten as
        \begin{widetext}
        \begin{align}
            \left( \partial^2_{T_0T_0}+2\epsilon \partial^2_{T_0T_1} \right. & \left. +\epsilon^2\partial^2_{T_1T_1} +\frac{R_{ms}}{M_{ms}\omega_\infty}\partial_{T_0}+\frac{R_{ms}}{M_{ms}\omega_\infty}\epsilon\partial_{T_1} + 1\right)\left(v_{0}(T_0,T_1) +\epsilon v_{1}(T_0,T_1)\right) \nonumber\\ &=  \epsilon \frac{S_d}{M_{ms}\omega_m}(\partial_{T_0}+\epsilon\partial_{T_1})\left[1-\left(1-\frac{Z_{ms}(\omega_c)}{\Tilde{Z}_t\left(1+A\cos\left(T_1 +\phi_m\right)\right)}\right)W(\omega)\right] p_f(T_0).
            \end{align}
            Separating the different orders in $\epsilon$, we get
            \begin{align}
          \mathcal{O}(\epsilon^0)  \rightarrow \left( \partial^2_{T_0T_0}+ \frac{R_{ms}}{M_{ms}\omega_\infty}\partial_{T_0}+1\right) v_0  = & 0,\label{eq:gov_v0}\\
          \mathcal{O}(\epsilon^1)  \rightarrow \left( \partial^2_{T_0T_0}+ \frac{R_{ms}}{M_{ms}\omega_\infty}\partial_{T_0}+1\right) v_1  =& \frac{S_d}{M_{ms}\omega_m}\partial_{T_0}\left[1-\left(1-\frac{Z_{ms}(\omega_c)}{Z_{t}\left(1+A\cos\left(T_1 +\phi_m\right)\right)}\right)W(\omega)\right] p_f(T_0)\nonumber\\&-(2 \partial^2_{T_0T_1}+\frac{R_{ms}}{M_{ms}\omega_\infty}\partial_{T_1})v_0. \label{eq:gov_v1}
        \end{align}

        \end{widetext}
        To solve the governing equation of the loudspeaker, we now have to solve each of the two separated partial differential equations giving the solution at order 0 and 1, $v_0$ and $v_1$ respectively.
        We first solve the partial differential equation at order $0$, eq.~\eqref{eq:gov_v0}, using the initial conditions. We then reinject the solution $v_0$ into eq.~\eqref{eq:gov_v1} and cancel the secular terms, to obtain the solution $v_1$. 
        
        Finally, using the definition of the velocity field expansion eq.~\eqref{eq:exp_v}, remembering that the two time scales are $T_0\approx \tau$, $T_1 \approx \epsilon \tau$, and replacing $\epsilon$ and $\tau$ by their definitions, we can derive the total solution to the initial governing equation, eq.~\eqref{eq:ode_speak}
        
        \begin{align}
               &\resizebox{.85\hsize}{!}{$v(t)  =  \frac{P_{inc}}{S_d Z_{ms}}\textrm{e}^{-\frac{R_{ms}}{2M_{ms}}t}\cos\left(\sqrt{1-(\frac{R_{ms}}{2M_{ms}\omega_\infty})^2}\omega_\infty t\right)$}  \\& \resizebox{.90\hsize}{!}{$ +\frac{P_{inc}}{Z_{ms}} \left[1-\left(1-\frac{Z_{ms}(\omega_c)}{Z_{t}\left(1+A\cos\left(\omega_m t +\phi_m\right)\right)}\right)W(\omega)\right]  \textrm{e}^{\textrm{i}\omega t}$} \nonumber.
        \end{align}     
        
        The first term corresponds to the transient field and decays exponentially with the speaker dissipation constant $R_{ms}/2M_{ms}$.
        At the control frequency, the loudspeaker responds effectively to the target impedance, $v(t) = P_{inc}/\left(Z_{t}\left(1+A\cos\left(\omega_m t +\phi_m\right)\right)\right)$. A constant fitting parameter is introduced such that $A=A_m/2$

        To extract the impedance matrix $\mathbf{Z}$ from the analytical model, we have to use the superposition principle and solve the system for an incident pressure at $\omega$, $\omega-\omega_m$, and $\omega+\omega_m$, and for $2N+1$ charges. Then, the Fourier transform of the pressure and velocity fields allows to solve eq.~\eqref{eq:Z_from_p_V} and thus to derive the scattering matrix eq.~\eqref{eq:Smatrix}. 

     \subsection{Numerical FDTD model}
        Numerical results are obtained using SIMULINK modelling of the entire experimental setup, based on a finite difference time step approach using a time step of $1.10^{-5}$. The scattering and impedance matrices can then be extracted from the incident and reflected pressure, using the 2-probe multi-load technique applied to the numerical experiment. The results obtained via the extraction procedure performed as in the experimental set-up are consistent with those obtained from the direct access to the reflected and transmitted pressures allowed by the simulation.

    \subsection{Numerical FEM model}    
    The numerical FEM experiment is performed using the frequency domain solver  of the commercial software COMSOL Multiphysics, following the methodology proposed in \cite{fleury_subwavelength_2015}. The time-modulated loudspeaker is modeled as an impedance
    \begin{align}
        Z &= Z_{ms}(\omega)+A_m[\textrm{i}\omega C_{ms}]^{-1}/S_d \cos(\textrm{i}\omega_m t+\phi_m)\\
        &=Z_{ms}(\omega)+\delta Z(\omega)(e^{\textrm{i}\omega_m t}e^{\textrm{i}\phi_m}+e^{\textrm{-i}\omega_m t}e^{\textrm{-i}\phi_m}).
    \end{align}
    
    An impedance condition is implemented as follows
    \begin{equation}
        -\mathbf{n}\frac{\mathbf{\nabla} p}{\rho_0} = p_f \frac{-\textrm{i}\omega}{Z}=-\mathbf{v}.\mathbf{n}\textrm{i}\omega.
    \end{equation}

    Expanding $p_f$ and $v$ (normal particle velocity) in Fourier series, we end up after some algebra to 
    \begin{align}
        p_{f_n} & \resizebox{.78\hsize}{!}{$=Z(\omega_n)v_n + \delta Z(\omega_n)\left(v_{n-1}e^{-\textrm{i}\phi_m} + v_{n+1}e^{\textrm{i}\phi_m}+...\right),$} \\
        & \textnormal{or equivalently}  \nonumber \\
        v_n & \resizebox{.78\hsize}{!}{$=\frac{p_{f_n}-\delta(\omega_n)\left(v_{n-1}e^{-\textrm{i}\phi_m} + v_{n+1}e^{\textrm{i}\phi_m}+...\right)}{Z(\omega_n)}$},
        \label{weak}
    \end{align}
    where $\omega_n = \omega+n \omega_m$.

    We then put eq.~\eqref{weak} into weak forms and solve for $n = (-2,-1,0,1,2)$ simultaneously for any incident frequency $\omega$.

    \subsection{Characterization of the resonator to control}
    A fitting procedure of the transfer function of the loudspeaker terminated by an open circuit, a short circuit, or a load $R=100.8$ $\Omega$, allows to obtain the following Thiele and Small parameters \cite{small1972closed-box}: $Bl = 3.63$ T·m, $M_{ms}= 2.9$ g, $C_{ms}= 0.214$ mm/N, and $R_{ms}= 0.54$ N.s/m. The natural resonance of the speaker occurs close to $200$ Hz. We thus choose to apply our control around the resonance to take advantage of the stable response in this range

\section{\label{app:detail}Effect and limitation of the control characteristics}

    \subsection{Change of the control frequency}
\begin{figure}
            \centering\includegraphics{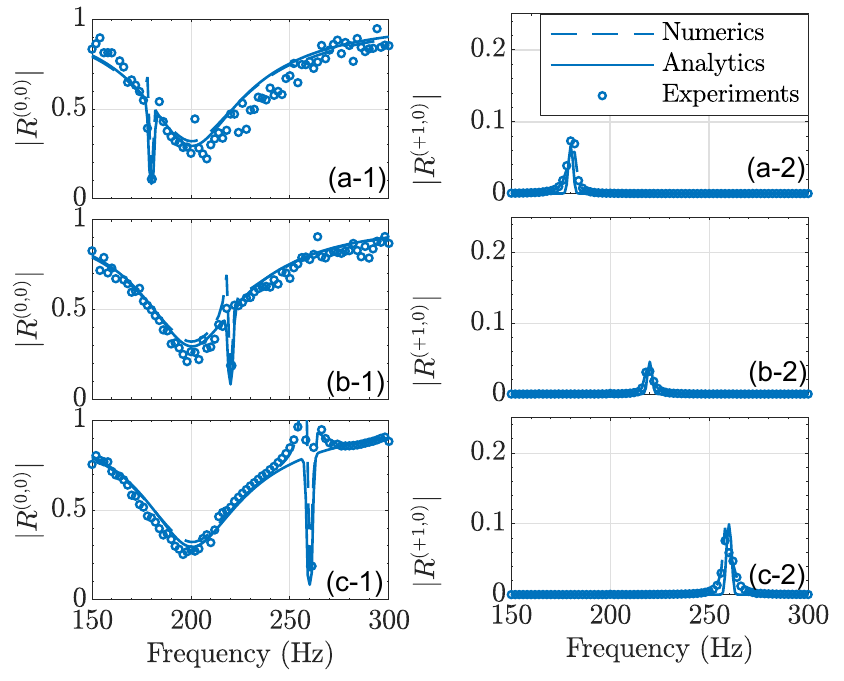}
            \caption{\textbf{Control frequency $f_c$ variation:} (a) $f_c = 180$ Hz, (b) $f_c = 220$ Hz, and (c) $f_c = 260$ Hz for the fundamental $|R^{(0,0)}|$ (1) and the 1st Floquet harmonic $|R^{(+1,0)}|$ (2). Analytical, numerical, and experimental results are given respectively by the solid and dashed lines and the symbols.}
            \label{fig:S_diff_fc}
        \end{figure}

        Figure~\ref{fig:S_diff_fc} shows three configurations with positive circular modulation and control frequency on either side of the natural resonance frequency, \textit{i.e.}, $f_c = 180$ Hz (a), $f_c = 220$ Hz (b) and $f_c = 260$ Hz (c).
        It can be seen that the active control does capture the reflection value $|R^{(0,0)}| = 0.11$ for each of the control frequencies and that a $+1$ Floquet harmonic is generated. The amplitude of the latter differs depending on the control frequency, again, due to the dispersion of the electrodynamic speaker.

    \subsection{Change of the target impedance $\Tilde{Z}_t$}
\begin{figure}
            \centering\includegraphics{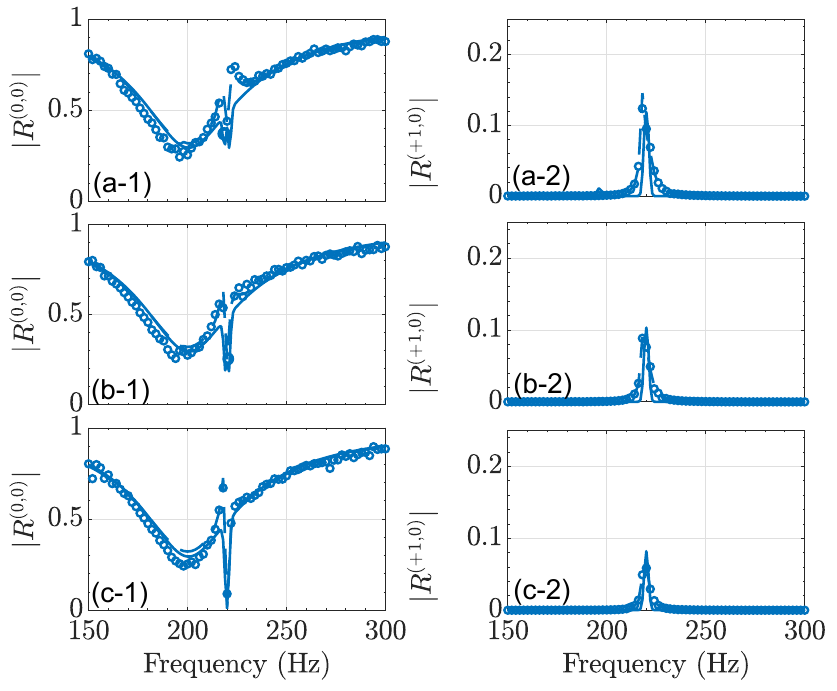}
            \caption{\textbf{Target impedance $\Tilde{Z}_t$ variation:} (a) $\Tilde{Z}_t= 0.4$, (b) $\Tilde{Z}_t = 0.6$, and (c) $\Tilde{Z}_t = 1$ for the fundamental $|R^{(0,0)}|$ (1) and the 1st Floquet harmonic $|R^{(+1,0)}|$ (2). Analytical, numerical, and experimental results are given respectively by the solid and dashed lines and the symbols.}
            \label{fig:S_diffZ}
        \end{figure}

        We then test in Fig.~\ref{fig:S_diffZ} three different target impedance values, respectively $\Tilde{Z}_t = 0.4$ (a), $\Tilde{Z}_t = 0.6$ (b), and $\Tilde{Z}_t = 1$ (c).
        Here again, the agreement between the three methods is rather good. The reflection at $\omega$ falls to $|R^{(0,0)}| = 0.28$, $|R^{(0,0)}| =0.25$, and $|R^{(0,0)}| =0$, respectively for $\Tilde{Z}_t=0.4$, $\Tilde{Z}_t=0.6$, and $\Tilde{Z}_t = 1$. The smaller the impedance, the larger the reflection at both $\omega$ and $\omega+\omega_m$. Furthermore, it should be noted here that even though for $\Tilde{Z}_t = 1$, we have an impedance matching and thus $|R^{(0,0)}| = 0$, a reflection at $\omega+\omega_m$ exists, $|R^{(+1,0)}| \neq 0$. The termination load is only impedance matched at the control frequency $f_c$ but not at $f_c+f_m$, in other words, $\mathbf{\Tilde{Z}_t}(1,2) \neq Z_0$.
        A vigilance point is that the control may be limited by instabilities if the loudspeaker is asked to respond with an impedance too different from its natural impedance at a given frequency.
        
       \subsection{Change of the modulation depth $A_m$}
\begin{figure}
            \centering\includegraphics{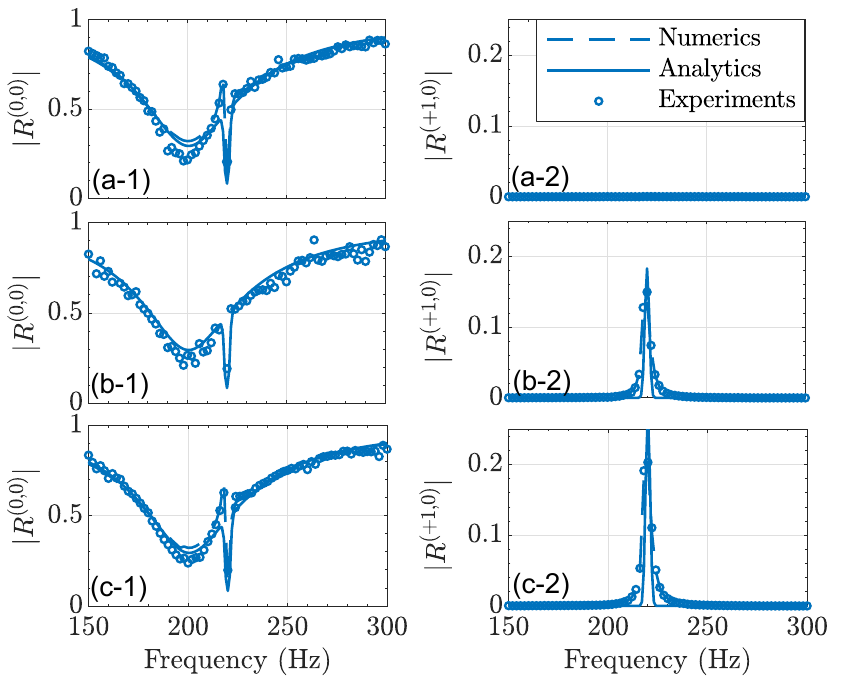}
            \caption{\textbf{Modulation depth $A_m$ variation:} (a) $A_m = 0$, (b) $A_m = 0.4$, and (c) $A_m = 0.6$ for the fundamental $|R^{(0,0)}|$ (1) and the 1st Floquet harmonic $|R^{(+1,0)}|$ (2). Analytical, numerical, and experimental results are given respectively by the solid and dashed lines and the symbols.}
            \label{fig:S_diffAM}
        \end{figure}

        Finally, we study the effect of changing the modulation amplitude, also called modulation depth. We assign three different values $A_m = 0$, $A_m = 0.4$, and $A_m =0.6$ as illustrated in Figs.~\ref{fig:S_diffAM}(a,b,c) respectively.
        
        As anticipated, the modulation depth has almost no impact on the reflection at the fundamental frequency. For zero modulation depth, Fig.~\ref{fig:S_diffAM}(a), \textit{i.e.}, a time-invariant control, no Floquet harmonics are generated. Furthermore, the higher the modulation depth, the higher the magnitude of the reflection at the 1st Floquet harmonic $|R^{(+1,0)}|$. A point of caution with the variation of $A_m$: a high modulation depth can also generate higher order harmonics. It is therefore important to check the magnitude of the reflection at $\omega+n\omega_m$, and if necessary, adapt the dimension of the $\mathbf{S}$ matrix to account for the additional harmonics in the calculation.
        \vspace{1cm}

\section{\label{app:details_expe_setup}Details on the experimental set-up}
   The acoustic apparatus used for the characterization of $\mathbf{S}$ and illustrated in Fig.~\ref{fig:exp_setup} consists of an acoustic waveguide made up of removable portions of a 7.18 cm diameter circular duct, terminated on one side by an electrodynamic loudspeaker (Monacor SPX-30 M, 3 inches) acting as the source, and on the other side, the time modulated load.  To consider only the propagation of plane waves, we take care to work only below the 1st cutoff frequency of the waveguide ($f_c = 1.8412 c_0/2\pi a = 1400$ Hz).

   The time-modulated load is an actively controlled electrodynamic loudspeaker (Monacor SPX-30 M, 3 inch) enclosed in a cavity of volume $V_b = 1081.6$ cm$^3$ and instrumented with an ICP microphone (PCB 130F20, 1/4 inch) placed just in front of the loudspeaker diaphragm. The active control scheme as well as the excitation signal generation and data acquisition are performed with an FPGA-based Speed-goat Performance real-time controller (I/O 131) controlled by the MATLAB/SIMULINK xPC target environment. The controller's output voltage is converted by a homemade voltage-to-current converter (0.2083 A/V) based on a Howland pump circuit and fed into the controlled speaker. 

   The incident and reflected pressure are derived from the pressure measured by two ICP microphones (PCB 130F20, 1/4 inch) 24 cm apart and 5 cm from the excitation source. To characterize the full scattering matrix, a multi-charge technique is used, with $L_{load}=50$, $62.5$, and $87.5$ cm respectively, thus ensuring a consistent difference in phase delay due to the round-trip propagation distance between the microphones and time-modulated load for each configuration. It is important to note that the distance should be chosen according to the center frequency of the bandwidth under consideration. Note that a further extension to multimode characterization is possible, but will require an increased number of sensors.


\end{document}